\documentclass[aps,prd, twocolumn,superscriptaddress,nofootinbib,preprintnumbers]{revtex4-1}
\usepackage{amsmath}
\usepackage{graphicx}
\usepackage{subfigure}
\usepackage{amssymb}
\usepackage{xcolor}
\usepackage{multirow}
\usepackage{cancel}
\usepackage{color}
\usepackage{ulem}
\usepackage{listings}
\usepackage{xcolor}
\usepackage{float}
\usepackage{slashed}
\usepackage{diagbox}

\usepackage[colorlinks,citecolor=blue]{hyperref}
\usepackage{amsmath}
\usepackage{wrapfig}

\usepackage[utf8]{inputenc} 

\newcommand{\be}{\begin{equation}}
\newcommand{\ee}{\end{equation}}
\newcommand{\bea}{\begin{eqnarray}}
\newcommand{\eea}{\end{eqnarray}}
\newcommand{\besp}{\begin{equation}\begin{split}}
\newcommand{\eesp}{\end{split}\end{equation}}

\newcommand{\nn}{\nonumber}

\newcommand{\Br}{\text{Br}}

\newcommand{\tabincell}[2]{\begin{tabular}{@{}#1@{}}#2\end{tabular}}
\newcommand{\Eq}[1]{Eq.~(\ref{#1})}

\newcommand{\Dfbd}{\mathord{\buildrel{\lower3pt\hbox{$\scriptscriptstyle\leftrightarrow$}}\over {D}_{\mu}}}

\def\mL{\mathcal{L}}

\def\mO{\mathcal{O}}

\def\1{\textbf{1}}
\def\2{\textbf{2}}
\def\3{\textbf{3}}
\def\4{\textbf{4}}
\def\5{\textbf{5}}
\def\6{\textbf{6}}
\def\7{\textbf{7}}
\def\8{\textbf{8}}
\def\9{\textbf{9}}

\begin{document}

\title{Probing Higgs exotic decay at the LHC with machine learning}

\author{Sunghoon Jung}
\email{sunghoonj@snu.ac.kr}
\affiliation{Center for Theoretical Physics, Department of Physics and Astronomy, Seoul National University, Seoul 08826, Korea}

\author{Zhen Liu}
\email{zliuphys@umn.edu}
\affiliation{School of Physics and Astronomy, University of Minnesota, Minneapolis, MN 55455, U.S.A.}

\author{Lian-Tao Wang}
\email{liantaow@uchicago.edu}
\affiliation{Enrico Fermi Institute, University of Chicago, Chicago, IL 60637, USA}
\affiliation{Department of Physics, University of Chicago, Chicago, IL 60637, USA}

\author{Ke-Pan Xie}
\email{kepan.xie@unl.edu}
\affiliation{Department of Physics and Astronomy, University of Nebraska, Lincoln, NE 68588, USA}
\affiliation{Center for Theoretical Physics, Department of Physics and Astronomy, Seoul National University, Seoul 08826, Korea}

\begin{abstract}

We study the tagging of Higgs exotic decay signals using different types of deep neural networks (DNNs), focusing on the $W^\pm h$ associated production channel followed by Higgs decaying into $n$ $b$-quarks with $n=4$, 6 and 8. All the Higgs decay products are collected into a fat-jet, to which we apply further selection using the DNNs.  Three kinds of DNNs are considered, namely convolutional neural network (CNN), recursive neural network (RecNN) and particle flow network (PFN). The PFN can achieve the best performance because its structure allows enfolding more information in addition to the four-momentums of the jet constituents, such as particle ID and tracks parameters. Using the PFN as an example, we verify that it can serve as an efficient tagger even though it is trained on a different event topology with different $b$-multiplicity from the actual signal. The projected sensitivity to the branching ratio of Higgs decaying into $n$ $b$-quarks at the HL-LHC are 10\%, 3\% and 1\%, for $n=4$, 6 and 8, respectively.

\end{abstract}

\maketitle

\newpage
\section{Introduction}

Higgs exotic decay is a promising window in probing new physics. The Higgs portal can provide the most relevant coupling between the Higgs boson and new physics, while its narrow Standard Model width enhances the sensitivity to exotic decay modes. Information on the exotic decay extracted from measurements of Higgs properties~\cite{CMS:2018uag,ATLAS:2019nkf,Cepeda:2019klc}, while useful, is less sensitive. At the same time, exclusive searches targeting specific channels are less efficient in casting a wide net. It would be useful to develop strategies that can bridge these two extremes.

As a benchmark scenario, we consider a dark sector consists of multiple dark scalars \cite{Strassler:2006im,Han:2007ae,Draper:2010ew,Curtin:2013fra,Craig:2015pha,Arkani-Hamed:2016rle,Meade:2018saz,Baldes:2018nel,Glioti:2018roy,Arkani-Hamed:2020yna,Carena:2021onl,Knapen:2021eip}. Due to their mixing with the Higgs boson, the final decay products would be heavy fermions, such as $b$-jets. At the same time, there could be cascade decays among the dark scalars, resulting in a variety of final states with different $b$-multiplicities. As such, this furnishes a good example in which a more universal ``tagger'' (rather than focusing on a particular final state) could be beneficial.

It is challenging to develop such a strategy. Searches based on simple cuts runs the risk of being too tailored to a specific feature or being too inclusive (hence less sensitive). It is a place where some of the machine learning techniques can shine. In this paper, we test several deep neural networks (DNNs) with the benchmark signal. These include convolutional neural network (CNN)~\cite{Cogan:2014oua,deOliveira:2015xxd,Baldi:2016fql,Barnard:2016qma,Kasieczka:2017nvn,Guo:2018hbv,Li:2019ufu}, recursive neural network (RecNN)~\cite{Louppe:2017ipp,Cheng:2017rdo,Fraser:2018ieu}, and particle flow network (PFN)~\cite{Komiske:2018cqr}.
See Refs.~\cite{Guest:2018yhq,Abdughani:2019wuv} and references therein for reviews of the DNN applications in LHC physics. In addition to comparing their performance on an individual channel, we also consider the universal applicability by testing them on the channels that they are not trained on.

The rest of this paper is organized as follows. Section~\ref{sec:benchmark} describes the details of the benchmarks as well as the preparation of machine learning data. Various DNNs are built and trained to distinguish the signals from backgrounds, as described in Section~\ref{sec:DNNs}. Their performances are then compared in Section~\ref{sec:performance}, where the current bound and projected reach of the exotic decay branching ratios are also given. We conclude in Section~\ref{sec:conclusion}.

\section{The benchmark processes}\label{sec:benchmark}

We consider the following Higgs exotic decay scenarios:
\begin{enumerate}
\item The $4b$ channel: $h\to a_0a_0$, $a_0\to b\bar b$;
\item The $6b$ channel: $h\to a_0a_0$, with one of the $a_0\to b\bar b$,  and the other $a_0\to a_1a_1$, followed by $a_1\to b\bar b$;
\item The $8b$ channel: $h\to a_0a_0$, with both $a_0\to a_1a_1$, followed by $a_1\to b\bar b$.
\end{enumerate}
\noindent where $a_0$ and $a_1$ are the new light neutral scalars, whose masses are chosen to be $M_0=30$ GeV and $M_1=12$ GeV respectively as benchmarks. For the production mechanism, we consider the $W^\pm h$ process with the leptonic decay $W^\pm\to\ell^\pm\nu$. The main background is the Standard Model (SM) $W^\pm+{\rm jets}$ with $W^\pm\to\ell^\pm\nu$, and the $t\bar t$ with semi-lepton decay. We consider the case in which all of the Higgs decay products are clustered into a fat-jet~\cite{Butterworth:2008iy}.

We write the model file of Higgs exotic decay with the {\tt FeynRules} package~\cite{Alloul:2013bka}. The parton-level events of signal and backgrounds at the 14 TeV LHC are simulated using the {\tt MadGraph5\_aMC@NLO}~\cite{Alwall:2014hca} package.
The events are matched to $+1{\rm~jet}$ final states and then interfaced to {\tt Pythia8}~\cite{Sjostrand:2014zea} for parton shower and hadronization, and to {\tt Delphes3}~\cite{deFavereau:2013fsa} for fast detector simulation. We use the CMS detector configurations as the detector setup.\footnote{We change the isolation $\Delta R$ parameters for electron and muon to 0.2 and 0.3, respectively, for increasing the signal acceptance with high multiplicity final states.}  We modify the $b$-tagging efficiency (and mis-tag rates for $c$-jet, light-flavor jets) to $0.77$ (and $1/6$, $1/134$) according to Ref.~\cite{Aaboud:2018xpj}.

\begin{table}
\footnotesize\centering\renewcommand\arraystretch{1.5}
\begin{tabular}{|c|c|c|c|c|c|c|}\hline
Cross section & \multicolumn{3}{c|}{Higgs exotic decay} & \multicolumn{3}{c|}{SM} \\ \cline{2-7}
[Unit: fb] & $\ell^\pm\nu4b$ & $\ell^\pm\nu6b$ & $\ell^\pm\nu8b$ & $W^\pm+{\rm jets}$ & $t\bar t$ & $W^\pm h$ \\ \hline
Boosted $\ell^\pm$ & 8.21 & 7.66 & 7.04 & $2.53\times10^5$ & $6.21\times10^3$ & 5.48 \\ \hline
fat-jet & 7.01 & 6.56 & 6.03 & $2.01\times10^5$ & $4.95\times10^3$ & 4.66 \\ \hline
$b$-veto & 6.17 & 5.80 & 5.35 & $1.96\times10^5$ & $2.17\times10^3$ & 4.07 \\ \hline
Mass window & 3.34 & 3.19 & 2.99 & $5.66\times10^3$ & 400 & 2.08 \\ \hline
Efficiency & 1.37\% & 1.36\% & 1.34\% & 0.96\% & 0.25\% & 1.31\% \\ \hline
\end{tabular}
\caption{The cut flow table for the pre-selection cuts before applying deep neutral networks. For the signals, the cross sections are given under the assumption of 100\% branching ratios, e.g. for $\ell^\pm\nu4b$ channel we assume $\Br(h\to4b)=100\%$. For the SM $W^\pm h$, we have used the branching ratio $\Br(h\to b\bar b)=58\%$~\cite{Tanabashi:2018oca}.}\label{tab:lv_events}
\end{table}

To suppress the background, we require the final state to have exactly one charged lepton with
\bea\label{l_cut}
&&p_T^\ell>25~{\rm GeV},\quad |\eta^\ell|<2.5, \nn \\
&&p_T^{\ell+\slashed{E}_T}>200~{\rm GeV},\quad M_T<100~{\rm GeV},
\eea
where the transverse mass is defined as
\be
M_T=\sqrt{2 p^\ell_T\slashed{E}_T\left(1-\cos\Delta\phi\right)},
\ee
with $\Delta\phi$ being the azimuthal angle difference between $\ell^\pm$ and $\slashed{E}_T$. We also demand at least one fat-jet reconstructed by the anti-$k_t$ algorithm with $\Delta R=1.5$ and
\be
200~{\rm GeV}<p_T^J<500 {\rm ~GeV}, \quad |\eta^J|<2.5~.
\ee
The fat-jets are trimmed by $R_{\rm cut}=0.3$ and $f_{\rm cut}=0.05$~\cite{Krohn:2009th}. Next, the small-$R$ jets are clustered using anti-$k_t$ algorithm with $R=0.4$, and $b$-tagged ones within
\be
p_T^b>25~{\rm GeV},\quad |\eta^b|<2.5~,
\ee
are vetoed to suppress the $t\bar t$ background. Finally, we require the mass of the leading fat-jet to be in the Higgs mass window
\be
100~{\rm GeV}<m_J<150~{\rm GeV},
\ee
and treat this jet as the Higgs-candidate. The cut flows for signals and backgrounds based on the leading-order (LO) cross sections are listed in Table~\ref{tab:lv_events}, where we also give the cross sections for the SM $W^\pm h\to\ell^\pm\nu b\bar b$.\footnote{To improve the event generation efficiency, without loss of generality we require at least one final state (light-) quark or gluon with $p_T>100$ GeV and 50 GeV for the $W^\pm+{\rm jets}$ and $t\bar t$ backgrounds at event generator level, respectively. These (light-) quark and gluons includes both ISR/FSR, as well as decay products from SM particles such as $W$ and $t$.} In particular, the composition of the data after pre-selection cuts in our sample is $W^\pm+{\rm jets}$ : $t\bar t$ : $W^\pm h =0.9337:0.0660:0.0003$. Beyond LO, the next-to-leading order (NLO) $K$-factor for the high-$p_T$ $W^\pm h$ production is $\sim1.5$~\cite{Ellis:1998fv,Campbell:2003hd,Butterworth:2008iy}. While for the main backgrounds, the $K$-factors for the SM $W^\pm+{\rm jets}$ and $t\bar t$ backgrounds are respectively $\sim1.5$ and $\sim1.6$ at next-to-next-to-leading order (NNLO)~\cite{Lindert:2017olm,Czakon:2011xx,Czakon:2013goa,Czakon:2012pz,Czakon:2012zr,Baernreuther:2012ws,Cacciari:2011hy}. Therefore, the LO-based analysis is not affected too much by the higher-order corrections.

\section{Building different DNNs}\label{sec:DNNs}

Around 30,000 events for each channel are generated, with 80\% and 20\% of the sample devoted to the training and validation/test datasets, respectively. We build DNNs to classify a specific exotic decay signal $W^\pm h\to\ell^\pm\nu nb$ (with $n=4$, 6 or 8) from the backgrounds. We focus on the $W^\pm +{\rm jets}$ and $t\bar t$ backgrounds for the study, and correspondingly the DNNs have three output neurons, one for the signal and the other two for backgrounds. The input features are basically the momenta of the constituents of the Higgs-candidate jet, but the concrete representation depends on the specific DNN we choose. We use three different types of DNNs, namely CNN, RecNN, and PFN, suitable for our purpose. These DNN constructions are described one by one in the following.

\subsection{CNN and jet images}

\begin{figure}
\centering
\subfigure{
\includegraphics[scale=0.28]{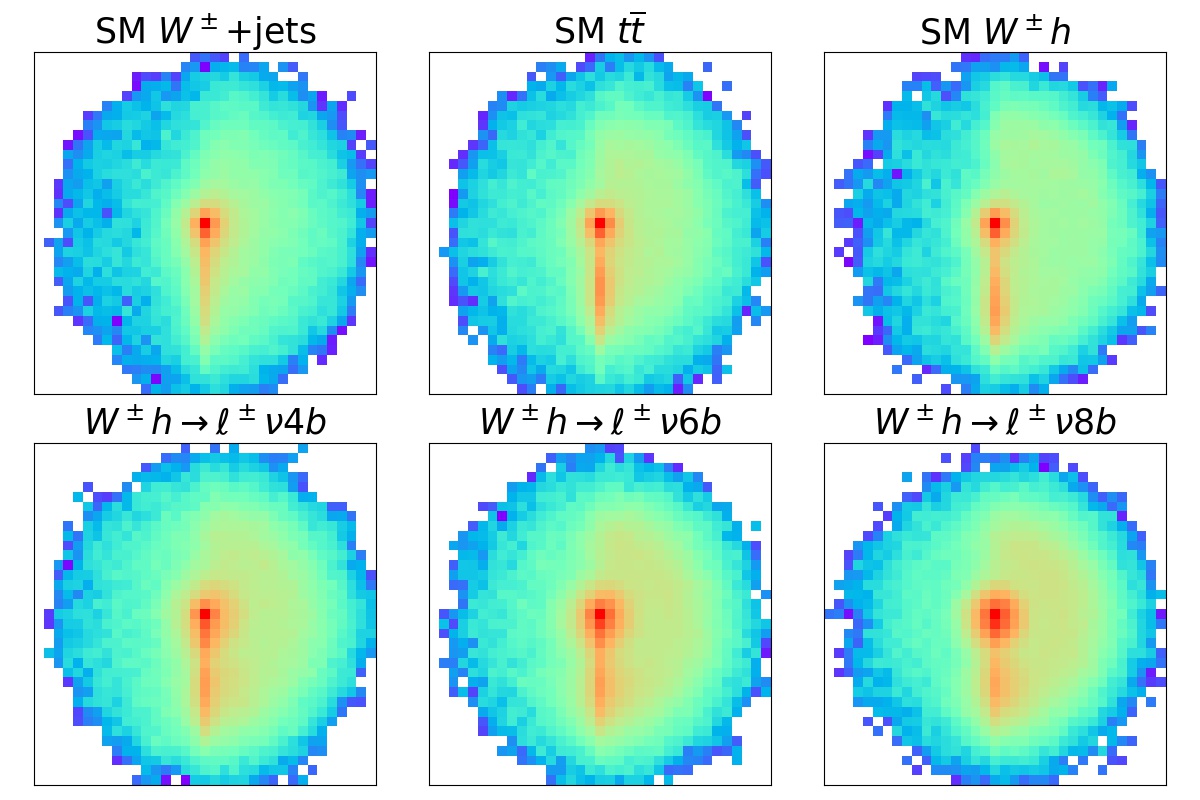}}
\caption{The average of 10,000 jet images for CNN. From blue to red, the color represents the increasing intensities of the pixels. }\label{fig:jet_images}
\end{figure}

We expand the constituents of the Higgs candidate in the $\eta$-$\phi$ plane to form a $35\times35$ pixels jet image with the granularity of $0.1\times0.1$. The intensity $I_i$ of a pixel $i$ is defined as the $p_T$ sum of all constituents inside it. We further normalize the intensities under the $L^2$-norm scheme so that $\sum_iI_i^2=1$~\cite{deOliveira:2015xxd}. To boost the learning efficiency, a series of pre-processing procedures are applied to the jet images: {\it i)} shifting the image so that the pixel with the highest intensity is in the center; {\it ii)} rotating the image so that the pixel with the second highest intensity is exactly below the center of the image; {\it iii)} reflecting the image so that the pixel with the third highest intensity is always lies in the right side. An individual jet image is rather sparse that only a few pixels are activated. To manifest the pattern of the images, in Fig.~\ref{fig:jet_images} we average 10,000 images for each channel. One can read from the figure that higher $b$-multiplicity gives more complex images for the exotic decay signals.

A CNN is built using the {\tt Keras}~\cite{chollet2015keras} package (with {\tt Tensorflow}~\cite{abadi2016tensorflow} as the backend) to classify the jet images. The 2-dimension information of an input image is first transformed into a 1-dimensional vector by a set of convolutional modules (denoted as CLs, whose content is a few {\tt Conv2D}, {\tt MaxPooling} and {\tt Dropout} layers), and then passed through the fully-connected layers (denoted as FLs, which consist of a few {\tt Dense} and {\tt Dropout} layers) to the output layer. The activation functions are {\tt ReLU} except for the output layer, where the {\tt softmax} function is used so that the three neuron outputs $r_{0,1,2}$ satisfy
\be
0<r_{0,1,2}<1,\quad r_0+r_1+r_2=1,
\ee
and hence can be interpreted as the probabilities of the signal and the two backgrounds, respectively. For a well-trained CNN, the output of signal events should form a peak around $r_0=1$.

We train one CNN for each signal channel. The {\tt Adam} optimization is adopted, and the batch size is set to be 1,000. The {\tt pool\_size} of {\tt MaxPooling} is chosen as $(2, 2)$, and step size of filtering is 1, and the parameter {\tt padding} is {\tt valid} which means the size of the image will reduce as the convolution applies. To choose the best configuration, we tune the hyper-parameters as follows.
\begin{enumerate}
\item For the CL module, we use 2 {\tt Conv2D} layers, with the filter numbers 32 or 64 per layer. The filter sizes are chosen as $[11,11]$, $[13,13]$, $[15,15]$, $[11,7]$ or $[15,13]$.
\item For the FL part, we try the structures of $[256,128]$, $[256,128,64]$ or $[512,256,128]$.
\item Different initial learning rates ($0.0001$, $0.0005$, $0.001$, $0.005$ or $0.01$) and dropout rates ($0.1$, $0.3$ or $0.5$) are also tested.
\end{enumerate}
The training is early-stopped if the accuracy of the validation data doesn't increase for ten epochs. This early-stopping can efficiently avoid overfitting.

\subsection{RecNN and natural languages}

The application of RecNN in classifying particle physics data is inspired by natural language processing (NLP). In this paradigm, the momenta are analogous to words, while clustering the constituents to reconstruct a jet is analogous to parsing a sentence~\cite{Louppe:2017ipp}. The jet clustering history is a full binary tree whose root is the reconstructed jet, while the leaves are the jet constituents.

The jet binary tree can be embedded into a vector as follows~\cite{Louppe:2017ipp}: first, for each node $k$ we attach a $q$-dimensional state vector $\textbf{u}_k$ by
\be
\textbf{u}_k=\sigma(W_u\textbf{v}_k+b_u),
\ee
where $\textbf{v}_k$ is the node's kinetic information (which we choose as the 7 observables $p$, $\eta$, $\theta$ $\phi$, $E$, $p_T$ and $E/E_J$), $W_u$ is a $q\times7$ weight matrix, $b_u$ is a $q\times1$ bias vector, and $\sigma$ is the {\tt ReLU} activation function. Second, for each node we further define a $q$-dimensional embedding vector $\textbf{h}_k$ recursively from the leaves to the root of the binary tree
\be
\textbf{h}_k=\begin{cases}~\textbf{u}_k,\quad&\text{if $k$ is a leaf;}\\ ~\sigma\left(W_h\begin{bmatrix}\textbf{h}_{k_L}\\ \textbf{h}_{k_R}\\ \textbf{u}_k\end{bmatrix}+b_h\right),\quad &\text{otherwise,}\end{cases}
\ee
where $W_h$ and $b_h$ are $q\times 3q$ weight matrix and $q\times1$ bias vector, respectively, while $h_{k_{L}}$ and $h_{k_{R}}$ respectively denote the embedded vector of the left- and right-children of node $k$ (where we require the left-child is more energetic by data pre-processing). Finally, we denote the embedding of the root as $\textbf{h}_{\rm jet}$, and interface it to a fully-connected module with two layers and $q$ hidden neurons per layer to classify the signal and backgrounds. The activation of the output layer is again {\tt softmax} for the purpose of a probability interpretation. Through this procedure, we have built up a RecNN with event-dependent tree structures, however they share the same weight matrices $W_u$, $W_h$ and bias vectors $b_u$, $b_h$, which will be optimized automatically when training the RecNN. On the other hand, the embedding dimension $q$ is the hyper-parameter that needs to be chosen before training.

\begin{figure}
\centering
\subfigure[~The clustering history of fat-jets from SM $W^\pm+{\rm jets}$.]{
\includegraphics[scale=0.085]{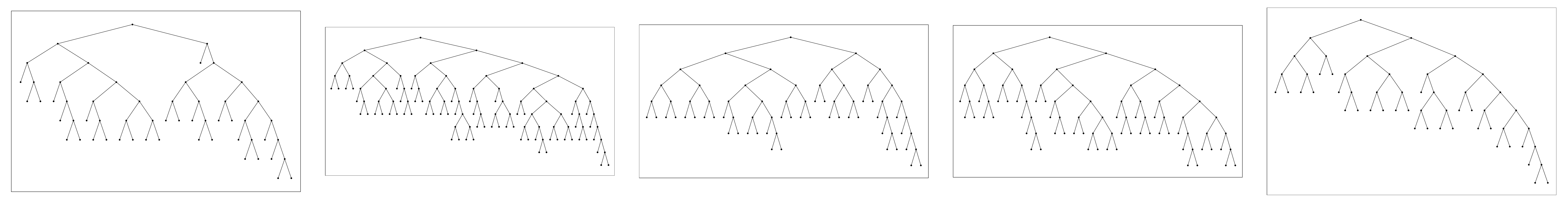}}
\subfigure[~The clustering history of fat-jets from SM $t\bar t$.]{
\includegraphics[scale=0.085]{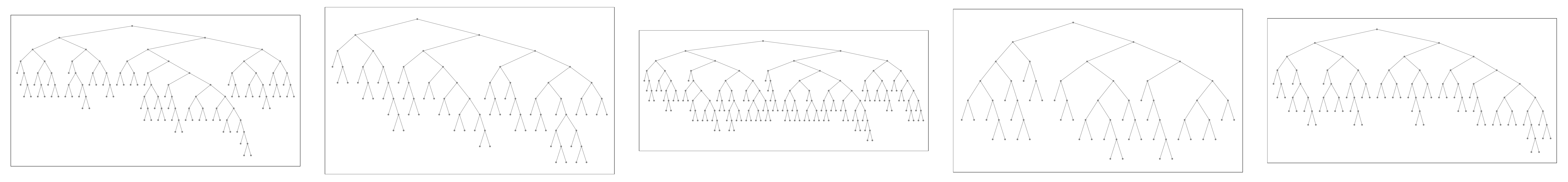}}
\subfigure[~The clustering history of fat-jets from SM $W^\pm h$.]{
\includegraphics[scale=0.085]{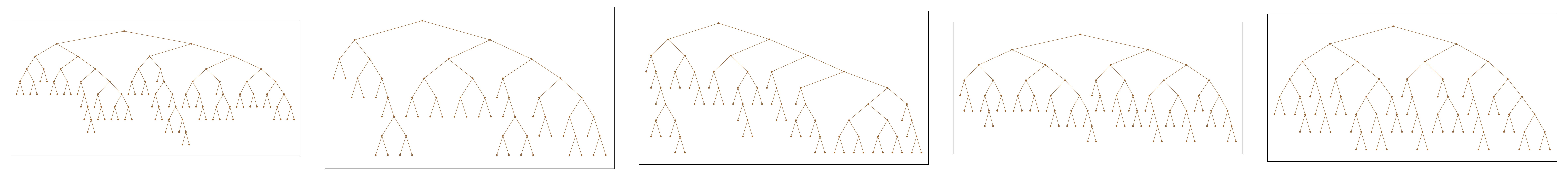}}
\subfigure[~The clustering history of fat-jets from $\ell^\pm\nu4b$.]{
\includegraphics[scale=0.085]{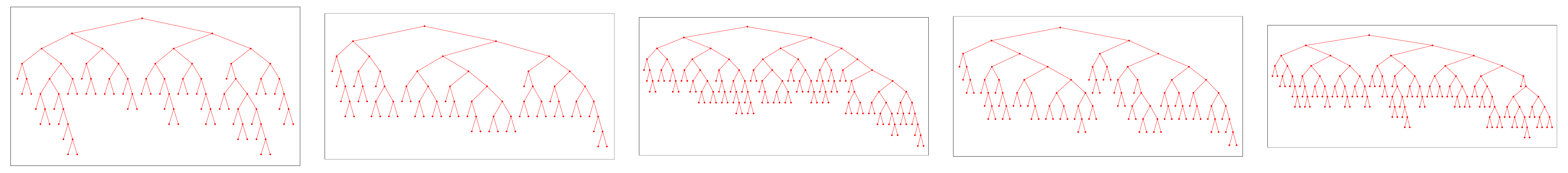}}
\subfigure[~The clustering history of fat-jets from $\ell^\pm\nu6b$.]{
\includegraphics[scale=0.085]{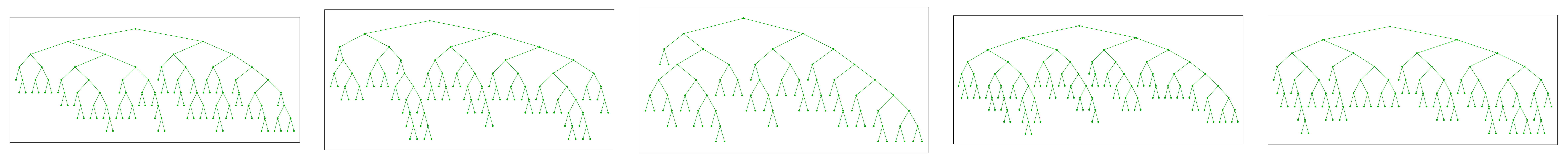}}
\subfigure[~The clustering history of fat-jets from $\ell^\pm\nu8b$.]{
\includegraphics[scale=0.085]{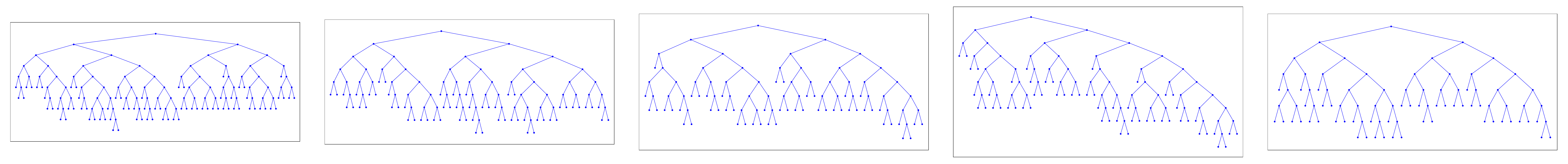}}
\caption{Illustrations of the jet clustering histories from signals and backgrounds for RecNN.}
\label{fig:cluster}
\end{figure}

As suggested by Ref.~\cite{Louppe:2017ipp}, we apply the $k_t$ algorithm to the constituents of the Higgs candidate for a second step reconstruction and get the clustering history. The illustrations of signals and backgrounds are shown in Fig.~\ref{fig:cluster}. The RecNN is built and implemented using the {\tt Python} codes provided in Ref.~\cite{Louppe:2017ipp}. To obtain the best configuration, we vary the embedding dimension $q$ in 40, 80, the initial learning rate in 0.001, 0.01, and batch size in 100, 1000, and compare the learning accuracies. As the training of RecNN is rather time-consuming, we stop the training when the accuracy on validation data does not increase for three epochs.

\subsection{PFN and tracks information}\label{sec:PFN}

Denoting the observables of a jet constituent as a $d$-dimensional vector $p=\{\xi_1,\xi_2,\cdots,\xi_d\}$, then most high-level jet observables (such as jet mass, multiplicity, track mass, momentum dispersion, etc) can be written in a form
\be\label{PFN_motivation}
\mO=F\left(\sum_i\Phi(p_i)\right),
\ee
where $\Phi(p)$ and $F(x)$ are respectively $\mathbb{R}^d\to\mathbb{R}^\ell$ and $\mathbb{R}^\ell\to\mathbb{R}$ functions determined by the observable $\mO$, while the summation index $i$ runs over all constituents of the jet. For example, for the jet mass $m_J$, we have $\Phi(p)\equiv p^\mu\equiv(E,p_x,p_y,p_z)$ and $F(x)=x^\mu x_\mu$. More examples can be found in Ref.~\cite{Komiske:2018cqr}. Inspired by this, Ref.~\cite{Komiske:2018cqr} proposes the PFN, which is based on \Eq{PFN_motivation}, but treats the $\Phi(p)$ and $F(x)$ as unknown functions to be constructed through the machine learning training process. More specifically, $\Phi(p)$ ($F(x)$) is represented by a set of fully-connected layers with $N_\Phi$ ($N_F$) hidden layers and neuron number $n_\Phi$ ($n_F$) per layer. We generally use the {\tt ReLU} function as activations, except for the output layer of $F(x)$ for which we use {\tt softmax}.

\begin{figure}
\centering
\subfigure{
\includegraphics[scale=0.285]{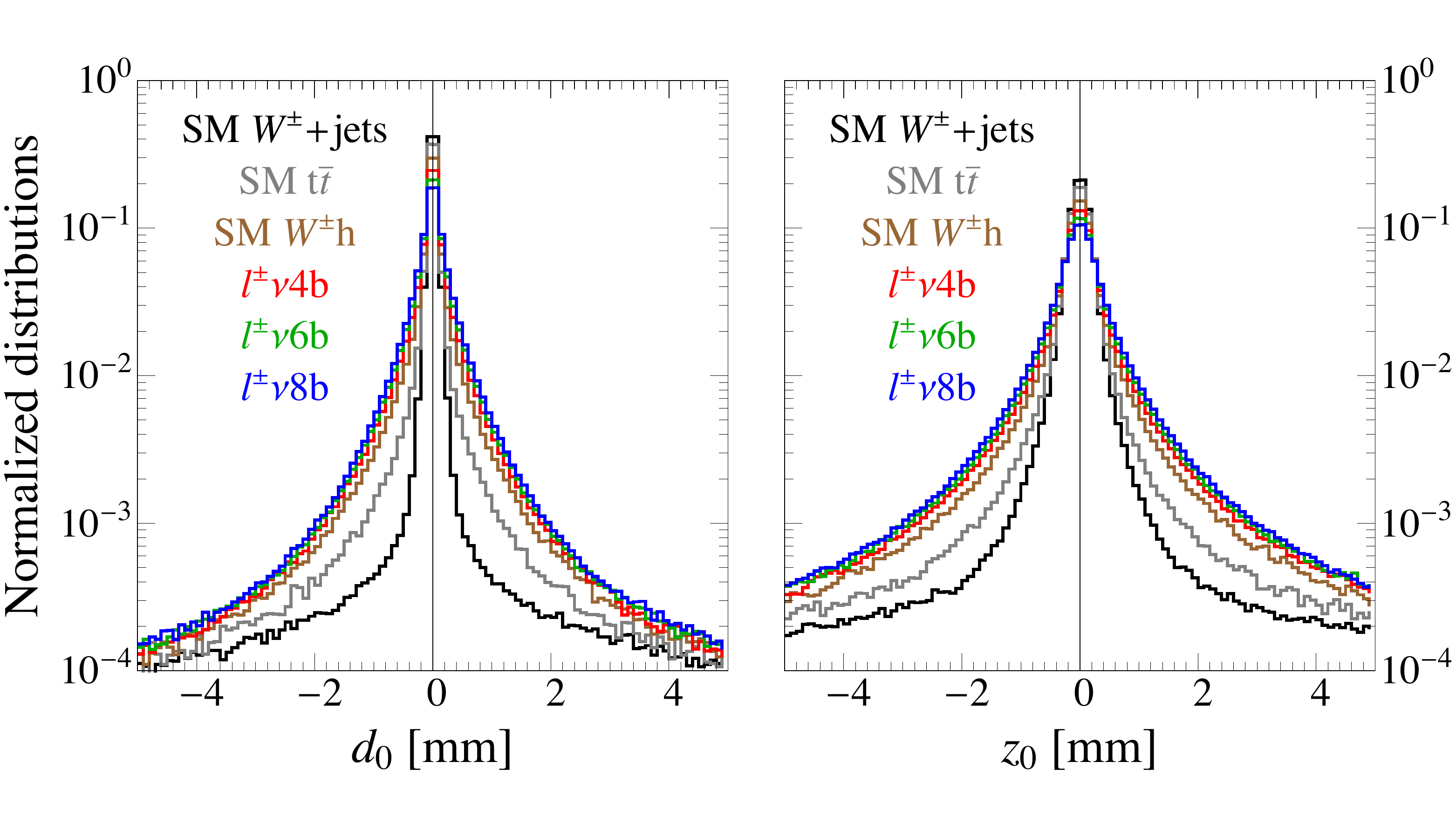}}
\caption{The impact parameter distributions $d_0$ and $z_0$ for the signals and backgrounds for PFN\_tracks.}
\label{fig:track}
\end{figure}

The ``primary'' setup for a PFN is to use the momenta of the jet constituents as input observables, i.e. $(p_T,\eta,\phi)$ and hence $d=3$ for a single particle. However, the advantage of PFN is that it can enfold extra information besides the four-momentum. For example, a jet constituent's particle ID (PID) can also be added as a component of $p$, and then $d=4$. Technically, we use the so-called ``float PID'' function mapping the PID into a real number between 0.1 and 1.1~\cite{Komiske:2018cqr}.\footnote{We only use different numbers to label charged particles such as electron, muon, proton, pion and kaon. The identification of charged particles can be done well, especially electron, muon and charged pions. We didn't model the difference between neutral hadrons such as $K_L^0$ and neutron.}
Such a PFN is denoted as PFN\_PID. Since the final state of the signal contains multiple $b$-jets, the tracks information could help distinguish them from the backgrounds. Therefore, we build an extended PFN (denoted as ``PFN\_tracks'') with the track impact parameters $d_0$ and $z_0$ as two additional $p$-components and hence $d=6$. As shown in Fig.~\ref{fig:track}, the impact parameter distributions have a longer tail for the events with higher $b$-multiplicities. The smearing effect is important for the impact parameter reconstruction, and we use the smearing procedure suggested by Ref.~\cite{Wildauer:2224523}.

The PFN classifiers are built using the {\tt PyTorch} package~\cite{NEURIPS2019_9015}. For simplicity, we fix
\be
N_\Phi=2,\quad N_F=3, \quad n_\Phi=n_F=100,
\ee
as a benchmark, and vary the single-particle embedding dimension $\ell$ in 32, 64, 128, and the batch size in 100, 200 to find the best configurations. During training, if there are three continuous epochs that the accuracy of validation data does not increase, we stop the training to prevent overfitting.

\section{DNN results}\label{sec:performance}

\subsection{Performance of different DNNs}

After training, we apply the DNNs to the test data and get the neuron output distributions. Due to the {\tt softmax} function of the output layer, the 0th output neuron $r_0$ peaks around 1 for the signal process, and peaks around 0 for the two backgrounds. Adding a cut of $r_0>r_c$ helps to enhance the signal significance. Varying the cut threshold $r_c$ from 0 to 1, we can get the signal efficiency versus background rejection curves. While the signal efficiency is defined as
\be
\epsilon_S\equiv\frac{N_S^{(r_0>r_c)}}{N_S^{(\rm total)}},
\ee
the background efficiency is defined as the weighted sum of the two backgrounds
\be
\epsilon_B\equiv\frac{\sigma_{B1}N_{B1}^{(r_0>r_c)}+\sigma_{B2}N_{B2}^{(r_0>r_c)}}{\sigma_{B1}N_{B1}^{(\rm total)}+\sigma_{B2}N_{B2}^{(\rm total)}},
\ee
where $B1$ and $B2$ represent the $W^\pm+{\rm jets}$ and $t\bar t$ backgrounds, respectively. We have checked that adding SM $W^\pm h$ as an extra neuron output to the DNNs give almost no change to $\epsilon_B$.

\begin{figure}
\centering
\subfigure{
\includegraphics[width=\columnwidth]{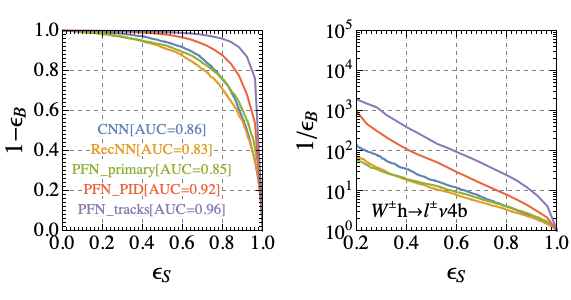}}
\subfigure{
\includegraphics[width=\columnwidth]{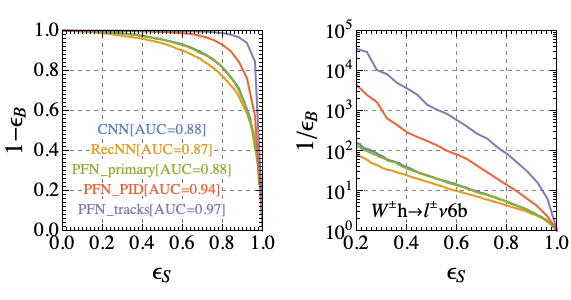}}
\subfigure{
\includegraphics[width=\columnwidth]{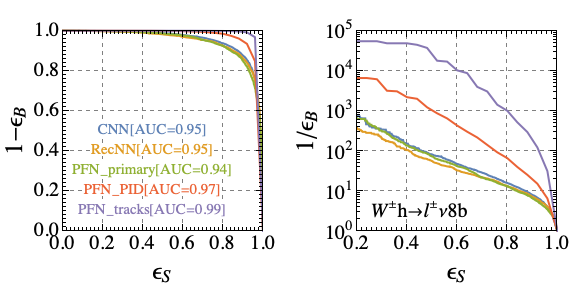}}
\caption{The signal efficiencies versus background rejections for the three signal benchmarks.}
\label{fig:ROC}
\end{figure}

The left panel of Fig.~\ref{fig:ROC} shows the $\epsilon_S$-$(1-\epsilon_B)$ curves, which are known as the receiver operating characteristic (ROC) curves. The area under curves (AUCs) are also shown in the figures. Another representation of the curves, i.e., the signal efficiency $\epsilon_S$ versus background rejection $1/\epsilon_B$, are shown in the right panel of Fig.~\ref{fig:ROC}. For the $\ell^\pm\nu6b$ and $\ell^\pm8b$ channels in the PFN\_tracks case, when the cut on $r_0$ is very close to 1, we might reach a background free region caused by limited statistics. In this case, we take a conservative estimate by replacing the background free region with a fixed sample event number $N_B^{\rm min}=0.1$.

From Fig.~\ref{fig:ROC}, we can conclude that CNN, RecNN, and PFN\_primary (i.e. with only four-momentum information of the jet constituents) can all serve as effective signal-selection and background-rejection taggers. Moreover, they have quite similar performances. For example, in the $4b$ channel, for a signal efficiency $\epsilon_S \sim 0.6$, we can have a rejection $\sim 10$. Since all DNNs with only constituents four-momentum inputs have similar performance, it is indicative of the fact that all the four-momentum information can be effectively and faithfully extracted by each one of them. For all the DNNs, the performance is better for higher $b$-multiplicities. For example, for the $\ell^\pm\nu8b$ sample, we can have a rejection of $40\div50$ for a signal efficiency $\epsilon_S \sim 0.6$. This is expected since the signal with higher $b$-multiplicities differs more from the major background from (sub-)jet multiplicities and distributions.

Adding more information about the signal indeed leads to better performances. As discussed in Section~\ref{sec:PFN}, we can incorporate the particle ID and track information into the PFN. The results are also shown in Fig.~\ref{fig:ROC}. A gain of about a factor of 10 is achievable by adding particle ID.  As one can expect, the displacement from the $b$-mesons imprints in the track impact parameters and can help improve the PFN performance.
The improvement with additional track information depends on the multiplicity of $b$-jets, with much better rejection for higher multiplicities (similar to the observation we made for other DNNs).

\begin{figure}[h!]
\centering
\subfigure{
\includegraphics[width=\columnwidth]{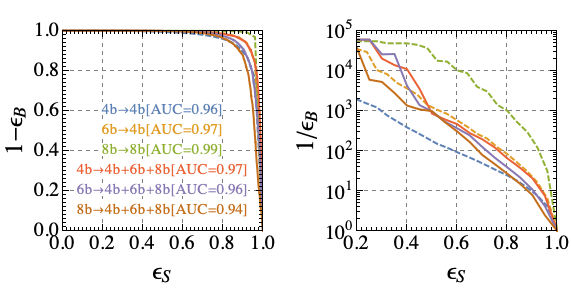}}
\caption{The universality of the PFN\_tracks. The notation ``$A\to B$'' in the figure means ``trained on process $A$ but tested on process $B$''.}
\label{fig:ROC_uni}
\end{figure}
In reality, Higgs exotic decay signals would be a combination of different processes and $b$-multiplicities. Without knowing the composition of the signal,  a more inclusive tagger for a class of final states. 
Motivated by this, we should aim for a tagger which can be applied efficiently even without being trained on the precisely the ``right" signal sample. To this end,  we further test the universality of the DNNs by applying a DNN trained on a specific signal channel to a mixed $4b+6b+8b$ sample with an equal amount of event numbers for each channel. To be concrete, we use PFN\_track for this test. The result of the signal efficiency on the mixed sample is shown in Fig.~\ref{fig:ROC_uni}. The DNNs trained on an exclusive sample still have a good performance on the mixed sample. For example, with $\epsilon_S \sim 0.6$, they have a rejection $1/\epsilon_B \sim 500$. The performance is better than that on the exclusive $\ell^\pm\nu4b$ sample while worse than that on the $\ell^\pm\nu8b$ sample. This result is expected due to the general improvement of the performance with the $b$-multiplicity observed earlier.
We also observe here the improvement associated with $b$-multiplicity for the DNNs trained on the ``wrong" sample. Moreover, DNNs trained with different exclusive samples have similar performances on the mixed sample.

\begin{table}[h!]
\scriptsize\centering\renewcommand\arraystretch{1.5}
\begin{tabular}{|c|c|c|c|c|c|c|c|c|c|c|c|c|}\hline
\multicolumn{2}{|c|}{~Classification accuracies~} & SM & \multicolumn{3}{c|}{\tabincell{c}{$M_0=30$ GeV\\$M_1=12$ GeV}} \\ \hline
\multicolumn{2}{|c|}{\diagbox{~Tested on}{Trained on~}} & $h\to b\bar b$ & $\ell^\pm\nu4b$ & $\ell^\pm\nu6b$ & $\ell^\pm\nu8b$ \\ \hline
SM & $h\to b\bar b$ & 67.1\% & 61.4\% & 58.1\% & 56.5\% \\ \hline
\multirow{4}{*}{\tabincell{c}{$M_0=30$ GeV\\$M_1=12$ GeV}} & $\ell^\pm\nu4b$ & 69.3\% & 73.1\% & 69.7\% & 68.1\% \\ \cline{2-6}
& $\ell^\pm\nu6b$ & 72.3\% & 77.0\% & 76.5\% & 74.9\% \\ \cline{2-6}
& $\ell^\pm\nu8b$ & 74.4\% & 79.4\% & 79.9\% & 79.4\% \\ \cline{2-6}
& $4b+6b+8b$ & $-$ & 76.4\% & 74.7\% & 73.6\% \\ \hline
\end{tabular}
\caption{Testing the universality of the PFN with tracks information.}\label{tab:universality}
\end{table}

In Table~\ref{tab:universality}, we show the classification accuracies of DNNs trained on an exclusive sample and applied to different samples, both exclusive and mixed. \footnote{The classification accuracy is defined as the ratio of ``correct predictions'' to the length of the test dataset, where the ``prediction'' for a given event is defined as the neuron with maximal output. For instance, we count this event's classification as a signal if the neurons have output with $r_0=0.5$, $r_1=0.1$ and $r_2=0.3$. By this definition, the accuracy of a random prediction is 33.3\% for a three-neuron output DNN.} It is interesting to note that the DNNs trained on lower $b$-multiplicity samples perform better when they are applied to higher multiplicity samples. For example, the DNN trained on $\ell^\pm\nu4b$ sample (with an accuracy of $73.1\%$) has an accuracy of $79.4\%$ on the $8b$ sample. Again, this observation implies the DNN trained in $4b$ samples relies on the $b$ (sub-)jet information. Note that the DNNs also tag the SM $h\rightarrow b\bar b$ events with $55\%\div61\%$ efficiency, implying that the $b$-jets and Higgs masses played important roles in the signal and background separation. When conducting a search for the exotic decays, one can apply other well-trained and optimized taggers for the $h\rightarrow b\bar b$ process, and hence we do not take this SM Higgs process as background when deriving the limits in the next subsection.

\subsection{Branching ratio upper limits for the exotic decay}

As an application of the techniques studied here, we derive a projection for the sensitivity to the Higgs exotic decays at the LHC and different future hadron colliders.

\begin{figure}[h!]
\centering
\subfigure{
\includegraphics[width=0.48\columnwidth]{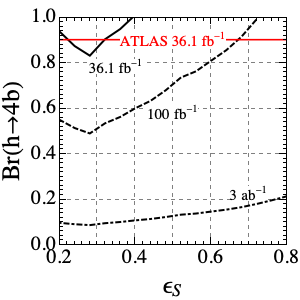}}
\subfigure{
\includegraphics[width=0.48\columnwidth]{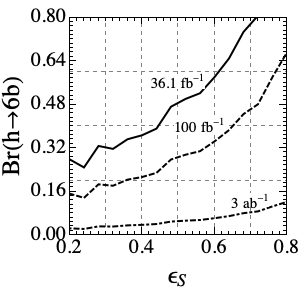}}
\subfigure{
\includegraphics[width=0.48\columnwidth]{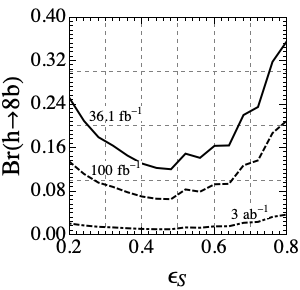}}
\caption{The branching ratio upper limits for the exotic decays at different luminosities, derived by the PFN\_tracks trained on the corresponding channel. For the $\ell^\pm\nu4b$ channel, the ATLAS result~\cite{Aaboud:2018iil} is also plotted as reference.}
\label{fig:brs}
\end{figure}

The projections are obtained as follows. Given a cut threshold $r_c$, we collect the event numbers of the signal and background samples that pass the cut, i.e. $N_S^{(r_0>r_c)}$ and $N_{B1}^{(r_0>r_c)}+N_{B2}^{(r_0>r_c)}$. They can be interpreted to the cross sections $\sigma_S$ and $\sigma_B$ after the cut. Therefore, given an integrated luminosity $\mL$ at the LHC, the corresponding expected signal and background event numbers.\footnote{To avoid confusion, we always use $N_{S,B}$ to denote the number of events in the training and validation/test samples, while use $S,B$ to denote the events normalized to a given integrated luminosity.}
\be\label{eq:sig}
S=\sigma_S\times\Br\times\mL,\quad B=\sigma_B\times\mL,
\ee
with $\Br$ being the branching ratio of the exotic decay channel. If no excess is obtained at the LHC, then signal event number upper limit $S_{\rm max}$ is determined by
\be\label{eq:Br}
\sqrt{2\left[S_{\rm max}-B\ln\left(1+\frac{S_{\max}}{B}\right)\right]}=2,
\ee
at 95\% confidence level. Combining \Eq{eq:sig} and \Eq{eq:Br} gives the branching ratio upper limit that can be achieved at a specific luminosity. The results for PFN\_tracks at different luminosities are shown in Fig.~\ref{fig:brs}. For the $\ell^\pm\nu4b$ channel, the ATLAS collaboration has a search on the non-boosted region at $36.1~{\rm fb}^{-1}$~\cite{Aaboud:2018iil},\footnote{Ref.~\cite{Aaboud:2018iil} studies both the $W^\pm h$ and the $Zh$ associated productions. For a proper comparison, here we only extract the single-lepton channel results.} which is also shown in the figure; our reach is comparable to the ATLAS. We can observe the maximal performance working point of signal efficiency around $0.2\div0.3$ for $4b$ and $6b$ samples, and around 0.45 for the $8b$ samples.

\begin{figure}[h!]
\centering
\subfigure{
\includegraphics[width=0.48\columnwidth]{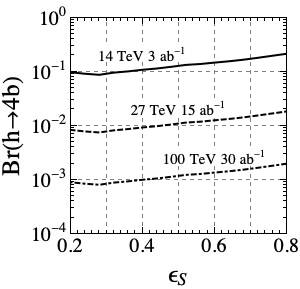}}
\subfigure{
\includegraphics[width=0.48\columnwidth]{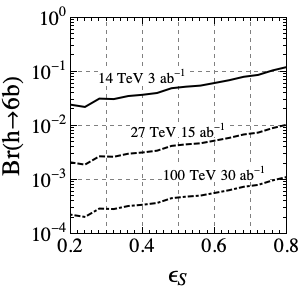}}
\subfigure{
\includegraphics[width=0.48\columnwidth]{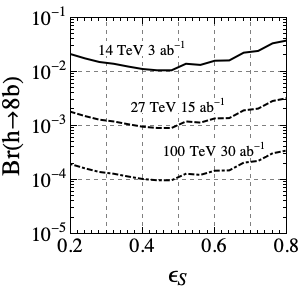}}
\caption{The branching ratio upper limits for the exotic decays at different colliders. Driven by naive rescaling of the 14 TeV results in PFN\_tracks.}
\label{fig:brs_rescaled}
\end{figure}
In Fig.~\ref{fig:brs_rescaled}, we have also shown the projection for a couple of proposed options of future hadron colliders, based on a simple rescaling via requiring the same signal events as in the HL-LHC. We can see that our semi-inclusive DNN taggers can help probe up to $10^{-4}$ of the Higgs exotic decays into multiple $b$-jets, comparable to what one can achieve at future lepton collider Higgs factories or lepton-hadron colliders with a much lower background~\cite{Liu:2016zki,Cheung:2020ndx}.

\section{Conclusion}\label{sec:conclusion}

In this article we study the possibility of probing the Higgs exotic decays via deep learning methods. We focus on the $W^\pm h$ associated production with Higgs decaying to $4b$, $6b$ or $8b$ final states. Such decays can go through a cascade of some intermediate scalars the intermediate light scalars. As a benchmark, we consider two such scalars $a_{0,1}$, for which the decay $a_0 \to a_1 a_1$ is possible. The decay products of the Higgs are collected into a fat-jet, whose kinetic information is fed to the DNNs for machine learning.

Comparing the results from CNN, RecNN and PFN\_primary, we found that although different DNNs use different representations to enfold the kinetic information, they yield very similar performances as long as the input observables include only the four-momentum of the jet constituents. This implies that each DNN has  efficiently extracted such information. In addition, by adding particle ID and track information as input observables, the PFN\_PID and PFN\_tracks can achieve a better performance. In addition, for a given type of DNN, signals with higher $b$-multiplicity have higher performances. We also test the universality of the DNNs by applying a PFN\_tracks model trained on a given signal channel (e.g. $\ell^\pm4b$) to another signal sample (e.g. $\ell^\pm6b$ or a mixed sample $4b+6b+8b$). Similar performances are found, implying that a well-trained tagger for a specific channel can also efficiently probe other exotic decay channels.

The training results from PFN\_tracks are used to derive the projected reach for the Higgs exotic decay branching ratios. We find that for the $4b$ channel, the reach is comparable to the existing ATLAS searches in the resolved (i.e. non-boosted) kinetic region. At the HL-LHC, $\Br(h\to4b)$ can be probed up to $\sim10\%$. For the $6b$ and $8b$ channels, the expected reach at the HL-LHC is respectively $\sim3\%$ and $\sim1\%$. By a simple rescaling, we also obtain the probe limits at the future colliders such as the 27 TeV HE-LHC and 100 TeV SppC/FCC-hh.

In order to make our result to be as independent of the production channel as possible, we have required the Higgs boson to have large boost, $p_T>200$ GeV, so that the decay products of the Higgs boson are relatively far away from other objects in the event. Hence, we expect the performance to be similar in other production channels. As an example, we have also tried the $Zh$ production with $Z\to\ell^+\ell^-$ and $\nu\bar\nu$, and obtained similar classification accuracies. Future work is needed to derive the precise performance in the other channels.

Our study showed that a semi-inclusive Higgs exotic decay DNN can be constructed with good performance. One can further explore such semi-inclusive BSM DNNs for other processes and decay modes. For instance, Higgs exotic decays into multiple scalars can also provide an admixture of $b$-jets and light jets (or kaons~\cite{CidVidal:2019urm}), and it would be useful to extend the search in this direction. Moreover, such a semi-inclusive approach particularly balances the broadness of the signal space and the usage of kinematics to suppress the background, which will complement the traditional exclusive searches and possible more model-agnostic searches. This strategy can be followed in many different channels, such as $W/Z/h$-decay to multiple sterile neutrinos~\cite{Chacko:2020zze}, continuum dark sector~\cite{Contino:2020god,Csaki:2021gfm,Csaki:2021xpy}, and generally hidden strong dynamics. It is also useful to extend the study of off-shell heavy standard model particle decays,  as well as the search of BSM particle decays such as $W^\prime$, $Z^\prime$, top partners and heavy Higgs.

\section*{Acknowledgement}

We thank Taoli Cheng, Yandong Liu, Gilles Louppe, Yongcheng Wu, Daneng Yang and Rui Zhang for useful discussions. We are especially grateful to Gilles Louppe for communication on RecNN and sharing the datasets. KPX would like to thank Guang-Ze Fu for the great help on coding.
The work of ZL is supported in part by the U.S.~Department of Energy under grant No. DE-SC0022345. SJ and KPX are supported by Grant Korea NRF-2019R1C1C1010050, and SJ also by POSCO Science Fellowship, and KPX also by the University of Nebraska-Lincoln. KPX would like to thank the hospitality of the University of Chicago where part of this work was performed. ZL and LTW acknowledge Aspen Center of Physics for hospitality during the final phase of this study, which is supported by National Science Foundation grant PHY-1607611.

\appendix
\section{Performance of the PFN on different mass benchmarks}

\begin{table*}
\scriptsize\centering\renewcommand\arraystretch{1.5}
\begin{tabular}{|c|c|c|c|c|c|c|c|c|c|c|c|c|c|c|c|}\hline
\multicolumn{2}{|c|}{~Classification accuracies~} & SM & \multicolumn{3}{c|}{\tabincell{c}{$M_0=30$ GeV\\$M_1=12$ GeV}} &\multicolumn{3}{c|}{\tabincell{c}{$M_0=50$ GeV\\$M_1=20$ GeV}} \\ \hline
\multicolumn{2}{|c|}{\diagbox{~Tested on}{Trained on~}} & $h\to b\bar b$ & $\ell^\pm\nu4b$ & $\ell^\pm\nu6b$ & $\ell^\pm\nu8b$ & $\ell^\pm\nu4b$ & $\ell^\pm\nu6b$ & $\ell^\pm\nu8b$ \\ \hline
SM & $h\to b\bar b$ & 67.1\% & 61.4\% & 58.1\% & 56.5\% & 56.3\% & 55.8\% & 52.3\% \\ \hline
\multirow{3}{*}{\tabincell{c}{$M_0=30$ GeV\\$M_1=12$ GeV}} & $\ell^\pm\nu4b$ & 69.3\% & 73.1\% & 69.7\% & 68.1\% & 67.4\% & 66.6\% & 62.0\% \\ \cline{2-9}
& $\ell^\pm\nu6b$ & 72.3\% & 77.0\% & 76.5\% & 74.9\% & 72.6\% & 73.1\% & 70.8\% \\ \cline{2-9}
& $\ell^\pm\nu8b$ & 74.4\% & 79.4\% & 79.9\% & 79.4\% & 76.1\% & 77.6\% & 77.0\% \\ \hline
\multirow{3}{*}{\tabincell{c}{$M_0=50$ GeV\\$M_1=20$ GeV}} & $\ell^\pm\nu4b$ & 62.4\% & 69.7\% & 67.1\% & 67.0\% & 72.9\% & 73.9\% & 70.7\% \\ \cline{2-9}
& $\ell^\pm\nu6b$ & 64.6\% & 73.9\% & 73.1\% & 73.1\% & 76.8\% & 77.3\% & 76.6\% \\ \cline{2-9}
& $\ell^\pm\nu8b$ & 66.5\% & 77.2\% & 76.6\% & 77.5\% & 79.4\% & 80.2\% & 80.1\% \\ \hline
\end{tabular}
\caption{The classification accuracies for two different mass benchmarks, extending Table~\ref{tab:universality}.}\label{tab:new_mass}
\end{table*}

To test the universality of our tagger at different mass choices, we setup another benchmark with $M_0=50$ GeV and $M_1=20$ GeV, and investigate the performance of PFN\_tracks. The results are shown in Table~\ref{tab:new_mass}. We can see that the DNN works well on this benchmark; moreover, the accuracies only decrease slightly when training on one mass benchmark and testing on another one. Therefore, a universal tagger with a weak dependence on the unknown masses of light scalars is plausible. One can explore this direction further.

\bibliographystyle{apsrev}
\bibliography{reference}

\end{document}